\documentclass[a4paper]{PoS}

\usepackage{graphics}
\usepackage{graphicx}
\usepackage{amsmath}
\usepackage{amssymb}
\usepackage{bm}
\usepackage{subfigure}

\newcommand{\tvec}[1]{\mathbf{#1}}

\newcommand{\ms}{\mskip 1.5mu}

\title{Double parton scattering in the ultraviolet: addressing the double counting problem}

\ShortTitle{DPS in the UV}

\author{Markus Diehl\\
        Deutsches Elektronen-Synchroton DESY \\
        Email: \email{markus.diehl@desy.de}}

\author{\speaker{Jonathan R.~Gaunt}\\
        Nikhef Theory Group and VU University Amsterdam\\
        E-mail: \email{jgaunt@nikhef.nl}}

\abstract{An important question in the theory of double parton scattering is how to incorporate 
the possibility of the parton pairs being generated perturbatively via $1 \to 2$ splitting into the 
theory, whilst avoiding double counting with single parton scattering loop corrections. Here, we 
describe a consistent approach for solving this problem, which retains the notion of double
parton distributions (DPDs) for individual hadrons. Further, we discuss the construction of 
appropriate model DPDs in our framework, and the use of these to compute the DPS part, presenting 
DPS `luminosities' from our model DPDs for a few sample cases.}

\FullConference{QCD Evolution 2016\\
		May 30-June 03, 2016\\
		National Institute for Subatomic Physics (Nikhef), Amsterdam\\
\vspace{10pt}
                $\textnormal{DESY 16-208}$}

\begin{document}

\section{Perturbative $1\to2$ splitting in DPS}

Whenever one has a final state in hadron-hadron collisions that can be split up into 
two subsets $A$ and $B$ with a hard scale in each (e.g. $WW$, $Wjj$, $4j$), the possibility exists for that final 
state to be produced in two separate hard collisions (double parton scattering, or DPS) 
rather than one (the more well-studied case of single parton scattering, or SPS). On the 
level of integrated cross sections, DPS is a power correction to SPS, but it is enhanced at 
small $x$ with respect to SPS (since it involves two parton ladders rather than one), and
can compete with SPS for certain processes where the SPS mechanism is suppressed by small
or multiple coupling constants (e.g. $W^\pm W^\pm$).

Earliest studies of DPS were conducted using the lowest order Feynman diagrams -- essentially
the parton model framework \cite{Paver:1982yp, Mekhfi:1983az} (see also \cite{Diehl:2011yj}). 
These studies indicated the following factorisation structure for this contribution:
\begin{align}
  \label{dps-Xsect}
 \frac{d\sigma_{\text{DPS}}}{dx_1\ms d\bar{x}_1\,
  dx_2\ms d\bar{x}_2}
&= \frac{1}{C}\; \hat{\sigma}_{ik\to A}\ms \hat{\sigma}_{jl\to B}
\int d^2\tvec{y}\;
F^{ij}(x_1, x_2, \tvec{y}) \, F^{kl}(\bar{x}_1, \bar{x}_2, \tvec{y}) \,,
\end{align}

Here, $\hat{\sigma}_{ij \to X}$ is the partonic cross section for the production of final state 
$X$ from partons $i$ and $j$, $C$
is a symmetry factor that is $2$ if $A=B$ and $1$ otherwise, and the $F^{ij}(x_1, x_2, \tvec{y})$
are the double parton distributions (DPDs). These depend on two $x$ fractions and flavours (for the two partons), 
as well as the quantity $\tvec{y}$ that measures the transverse separation between the two partons. This formula 
would then be simply added to the usual SPS cross section when computing the total cross section for production of $AB$.

\begin{figure}
\begin{center}
\subfigure[]{\includegraphics[height=4em]{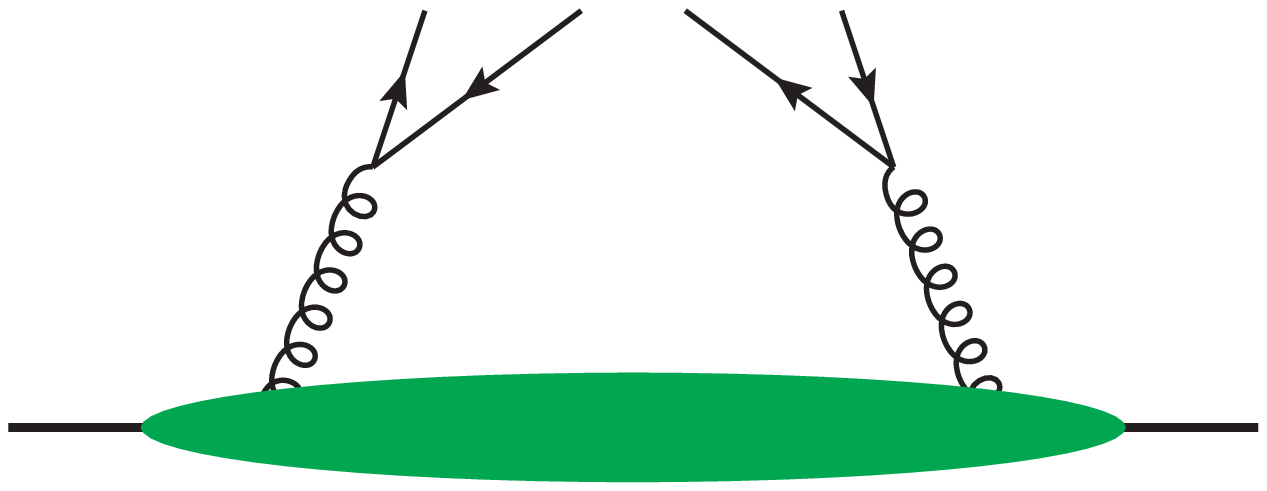}}
\hspace{1em}
\subfigure[]{\includegraphics[height=6em]{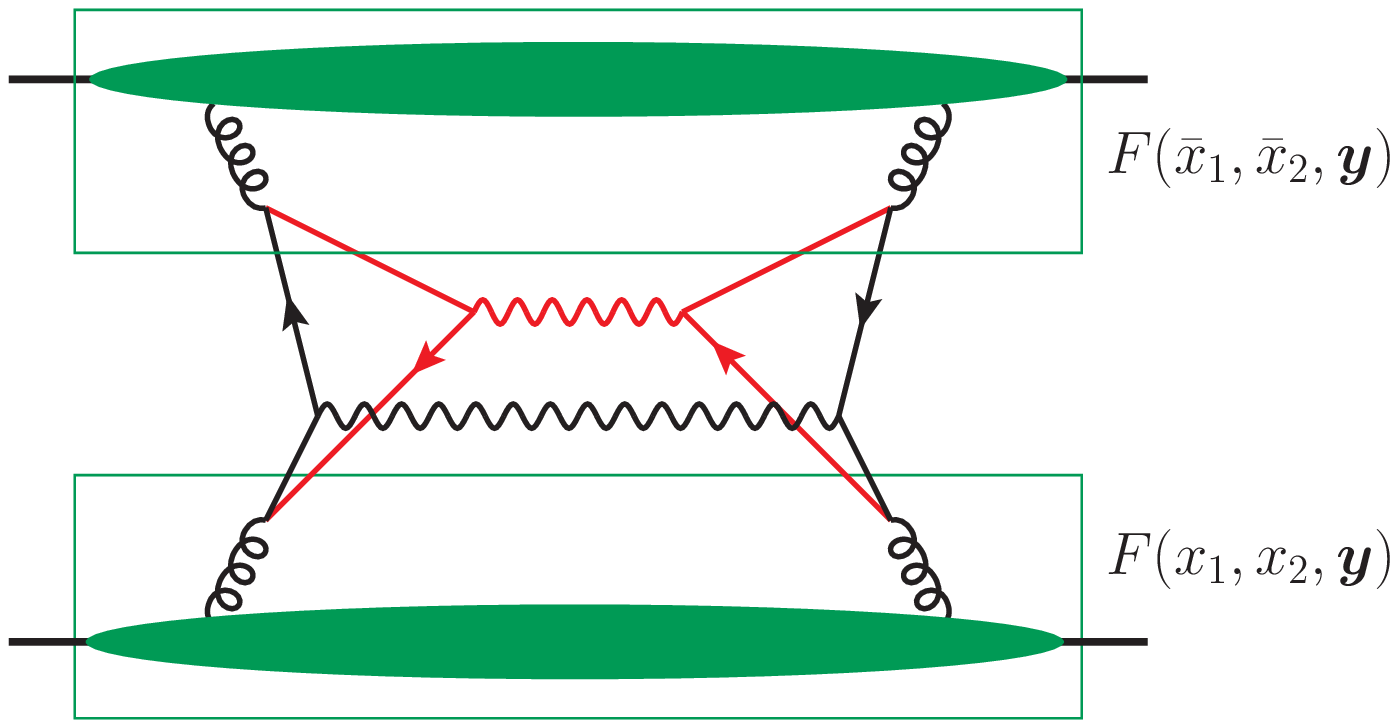}}
\hspace{1em}
\subfigure[]{\includegraphics[height=6em]{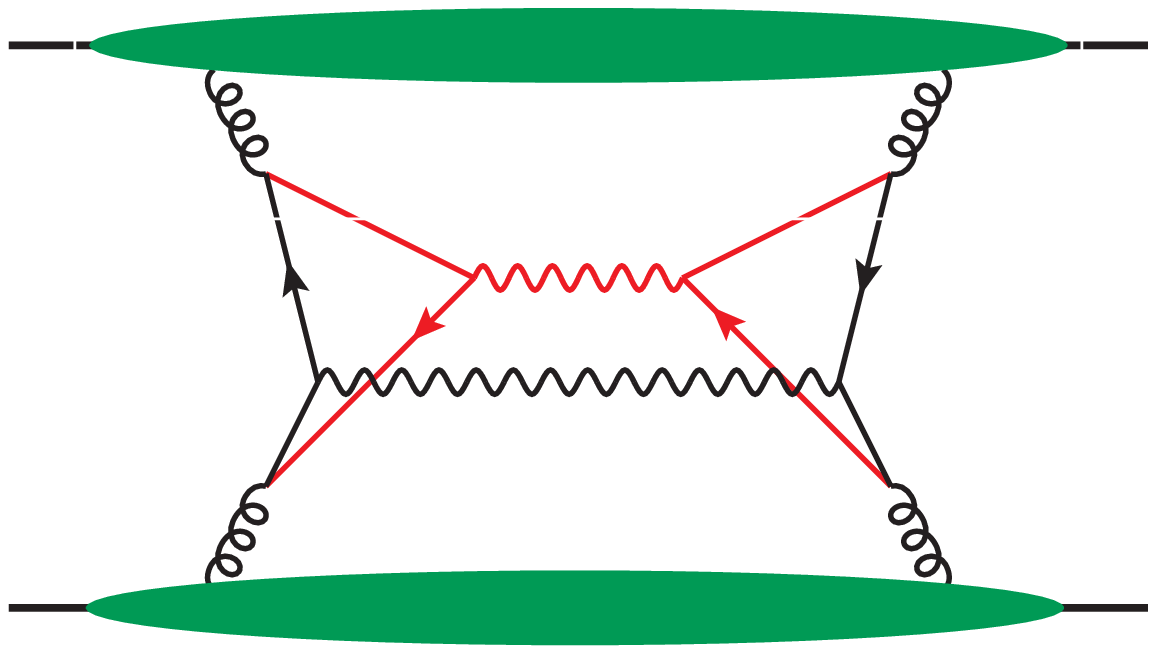}}
\caption{\label{fig:split} (a) Perturbative splitting contribution to a
  DPD.  (b) Contribution of double perturbative splitting to DPS, also
  called ``1v1'' graph.  (c) Single hard scattering contribution.}
\end{center}
\end{figure}

In recent years, efforts have been made to upgrade this picture to full QCD incorporating pQCD evolution
effects. Some of these effects are similar as are encountered in SPS -- i.e. diagonal parton emission from
one of the parton legs. These can be straightforwardly incorporated in a similar way as is done for SPS.
However, for DPS a new effect is possible -- as one goes backward from the hard interaction, one can find
that the DPS parton pair arose from a perturbative `$1\to2$' splitting (see figure \ref{fig:split}(a)). 
The perturbative splitting mechanism yields a contribution to the DPD of the following form:
\begin{align}
  \label{split-dpd}
F^{ij}(x_1,x_2, \tvec{y}) &= \frac{1}{y^2}\, \frac{\alpha_s}{2\pi^2}\,
\sum_k \frac{f_k(x_1+x_2)}{x_1+x_2}\, T_{k\to ij}\biggl( \frac{x_1}{x_1+x_2} \biggr) 
\end{align}
$f$ is the usual single PDF, $T$ is a splitting function, and $y \equiv |\tvec{y}|$. The $1/y^2$ 
behaviour of this contribution can be deduced already from dimensional counting grounds. 

Consistently incorporating the effects of $1\to2$ splittings in the theory is not straightforward. 
If one simply adds in a naive way the contribution of eq.~\eqref{split-dpd} to the DPD, the integral over 
$\tvec{y}$ in eq.~\eqref{dps-Xsect} becomes power divergent at small $y$. This power divergence 
appears in `1v1' diagrams with perturbative $1\to2$ splittings in both protons, as in figure 
\ref{fig:split}(b). Note that the graph in figure \ref{fig:split}(b) can also be viewed as a higher-loop 
correction to the leading power SPS process, as in figure \ref{fig:split}(c). This fact actually 
explains in an intuitive way the appearance of the power divergence in figure \ref{fig:split}(b) --
it comes from the 1v1 DPS diagram at small $y$ `leaking' into the higher-power SPS region.
The divergence at small $y$ is not present in reality (it arises from using DPS approximations 
in the small $y$ region where they are not valid) -- it should be removed, and replaced with
the appropriate SPS expression, in an appropriate way to avoid double counting between SPS and DPS.

\begin{figure}[t]
\begin{center}
\subfigure[]{\includegraphics[height=6em]{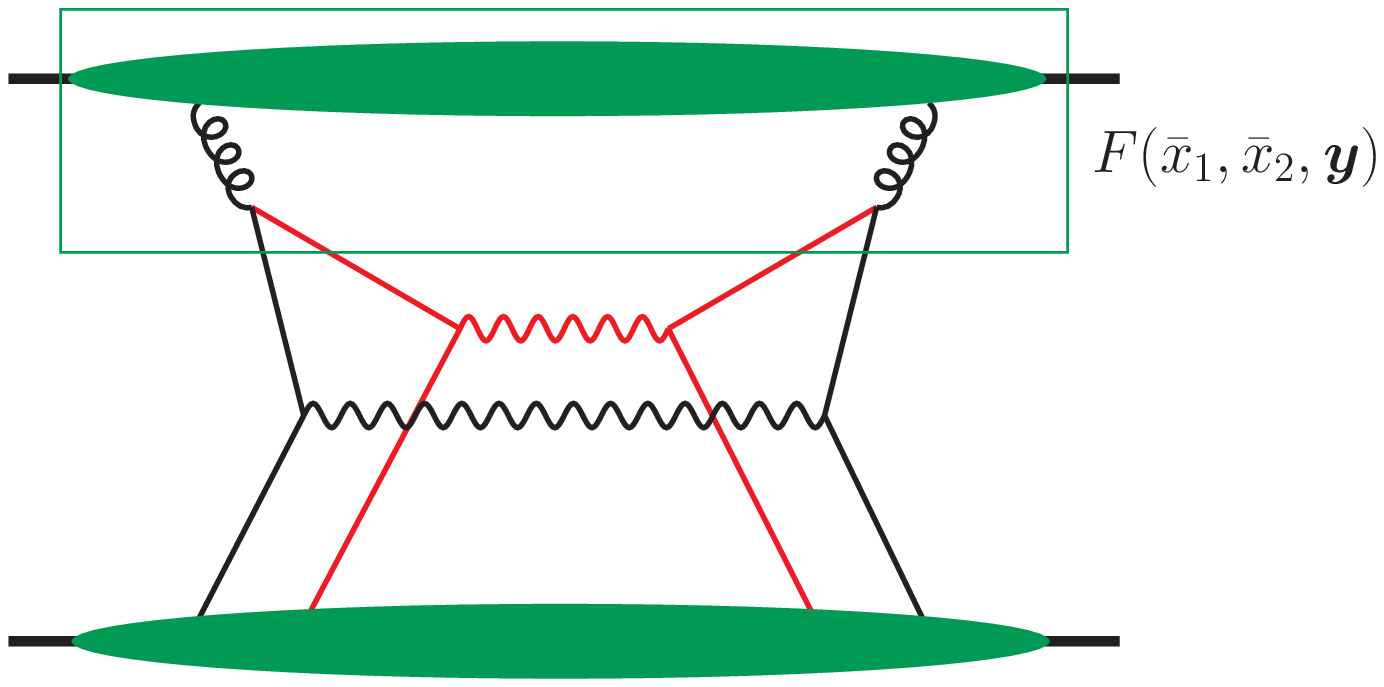}}
\hspace{1em}
\subfigure[]{\includegraphics[height=6em]{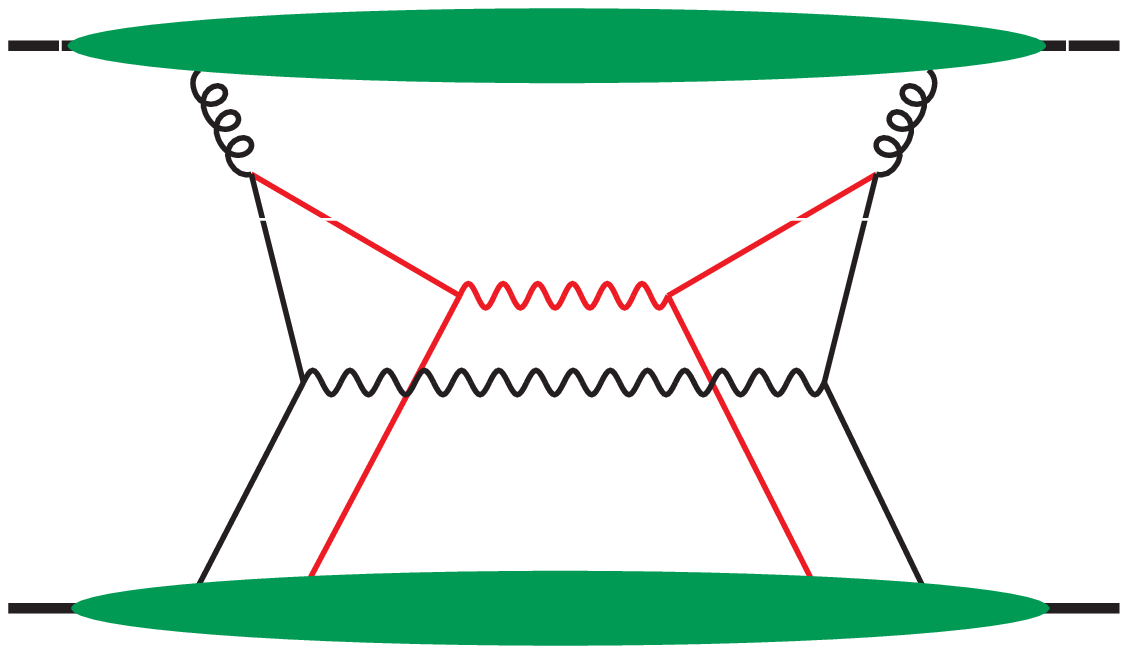}}
\caption{\label{fig:1vs2} (a) Contribution of single perturbative
  splitting to DPS, also called ``2v1'' graph.  (b) Graph with a
  twist-two distribution for one proton and a twist-four distribution for
  the other.}
\end{center}
\end{figure}

A divergence also appears in the $\tvec{y}$ integral for `2v1' diagrams with a $1\to2$ splitting in 
only one proton -- see figure \ref{fig:1vs2}(a). However, in this case we have only a logarithmic 
divergence, which one can associate with the overlap of the DPS contribution with the same-power
twist-two vs twist-four contribution (see figure \ref{fig:1vs2}(b)).   

In one previously-suggested approach for treating the $1\to 2$ splitting effects \cite{Blok:2011bu}, 
one makes a separation of the DPD into a `perturbative splitting' piece, and an `intrinsic' piece 
where the parton pair existed already at the nonperturbative scale. One includes the intrinsic 
$\otimes$ intrinsic (`2v2') and splitting $\otimes$ intrinsic (`2v1') contributions in DPS,
but simply discards the splitting $\otimes$ splitting (`1v1') contribution. This avoids double counting
with the SPS. The trouble with this approach is in the definition of `splitting' versus 
`intrinsic' pieces -- we do not know how such a separation could be achieved in a field 
theoretic definition valid at all $y$.

Another suggestion \cite{Manohar:2012pe} involves regulating the $y$ integral in eq.~\eqref{dps-Xsect} 
using dimensional regularisation. This also avoids double counting with the SPS, but a drawback
is that one loses the concept of the DPD of an individual hadron -- the appropriate operators 
for DPS then involve both hadrons at once.

A further past suggestion \cite{Ryskin:2011kk} is somewhat similar to \cite{Blok:2011bu}, but 
involves including the 1v1 contribution with an ad-hoc lower cut-off in the $y$ integral at 
values of order $1/Q$ (in \cite{Ryskin:2011kk} the cut-off is actually imposed in the Fourier
conjugate space, but the principle is the same). This renders the DPS contribution finite. However, 
there is now inevitably some double counting between DPS and SPS. There is in general a sizeable 
contribution to 1v1 DPS coming from the small $y$ region where the DPS picture is not valid. Finally,
there is a strong (quadratic) dependence of the DPS cross section on the unphysical cut-off -- 
adjusting the cut-off to other reasonable values such as $2/Q$ or $1/(2Q)$ will significantly affect
the cross section.
 
\section{A consistent scheme}

We now outline an alternative prescription which overcomes the drawbacks of the previous approaches.
First, we regulate the DPS cross section through the insertion of a cutoff function $\Phi(u)$ in 
the $\tvec{y}$ integral of eq.~\eqref{dps-Xsect}:
 \begin{align}
  \label{dps-reg}
\int d^2\tvec{y}\; \bigl[ \Phi(\nu y) \bigr]{}^2\,
  F^{ij}(x_1, x_2, \tvec{y}) \, F^{kl}(\bar{x}_1, \bar{x}_2, \tvec{y}) \,,
\end{align}
where $\Phi(u) \to 0$ for $u\to 0$ and $\Phi(u) \to 1$ for $u \gg 1$. This cuts out contributions
with $1/y$ much bigger than the cutoff scale $\nu$ from what we define as DPS, and regulates the
power divergence. An appropriate choice for this cutoff scale is $\nu \sim Q$. Note that here
$F^{ij}(x_1, x_2, \tvec{y})$ is the full DPD incorporating both parton pairs that had their origin in a 
perturbative $1\to2$ splitting, and those that did not. This enables us to define DPDs via operator matrix elements, without recourse to perturbation theory.

Thus far, the prescription resembles closely that of \cite{Ryskin:2011kk}, and suffers from double
counting between SPS and DPS. To fix this, we introduce a double counting subtraction term into the
total cross section formula including both SPS and DPS:
\begin{align}
  \label{full-Xsect}
\sigma_{\text{tot}} = \sigma_{\text{DPS}} - \sigma_{\text{sub}} + \sigma_{\text{SPS}} \,,
\end{align}
The subtraction term is given by the DPS cross section with both DPDs replaced by a
fixed order $1\to2$ splitting expression (at lowest order one simply has eq.~\eqref{split-dpd}
for each DPD) -- i.e. combining the approximations used to compute 1v1 splitting graphs 
in the two approaches. Note that at any order in $\alpha_s$, the computation of 
$\sigma_{\text{sub}}$ is technically much simpler than that of $\sigma_{\text{SPS}}$.

Let us demonstrate how this prescription works. At small $y$, of order $1/Q$, the dominant 
contribution to the DPD comes from the (fixed order) perturbative splitting expression 
(eq.~\eqref{split-dpd} at lowest order) -- thus one has $\sigma_{\text{DPS}} \simeq \sigma_{\text{sub}}$
and $\sigma_{\text{tot}} \simeq \sigma_{\text{SPS}}$ here as desired. The dependence on the
unphysical cut-off $\nu$ cancels between the subtraction and DPS terms. At large $y \gg 1/Q$,
the dominant contribution to $\sigma_{\text{SPS}}$ comes from the region of 1v1-type loops
where the DPS approximations are valid, such that $\sigma_{\text{SPS}} \simeq \sigma_{\text{sub}}$
and we have $\sigma_{\text{tot}} \simeq \sigma_{\text{DPS}}$ as appropriate. The construction just 
explained is a special case of the general subtraction formalism discussed in chapter 10 of 
\cite{Collins:2011zzd}, and it works order by order in perturbation theory.

So far we skirted over the issue of double counting between the 2v1 diagrams and the twist-two
vs. twist-four contributions. This can be fixed in an analogous way to the 1v1/SPS double counting,
yielding the following for the total cross section:
\begin{align}
  \label{all-Xsect}
\sigma_{\text{tot}} &= \sigma_{\text{DPS}}
   - \sigma_{\text{sub (1vs1)}} + \sigma_{\text{SPS}}
   - \sigma_{\text{sub (1vs2)}} + \sigma_{\text{tw2 $\times$ tw4}} \,.
\end{align}

One can show that the sum $(- \sigma_{\text{sub (1vs2)}} + \sigma_{\text{tw2 $\times$ tw4}})$ 
is subleading in logarithms $\log(Q/\Lambda)$ compared to the other terms (where $\Lambda$ is
an infrared scale), so can be dropped at leading logarithmic order.

Our formalism also appropriately resums DGLAP logarithms in the 1v1 and 2v1 diagrams in regimes
where this is appropriate, and can be extended in a straightforward way to the case of measured transverse momentum. We will not discuss these issues further here, referring the reader to \cite{Diehl:2016khr} for 
more detail.

\section{Double Parton Scattering Luminosities}

Here we make quantitative estimates of the DPS part of the cross section in our framework. 
In particular, we will present values for the $\tvec{y}$ integral in eq.~\eqref{dps-reg}, which 
we shall refer to as the DPS luminosity $\mathcal{L}$. We remind the reader that this is only part of the cross section for the production of $AB$, and can have a strong dependence on the cut-off parameter
$\nu$. We will discuss in these proceedings only the luminosity in the unpolarised case.

To make such estimates, one needs numerical values for the DPD $F^{ij}(x_1, x_2, \tvec{y},\mu)$
(we take the renormalisation scale for the two partons to be equal, and write this scale $\mu$
explicitly here). At perturbatively small $y \ll 1/\Lambda$, the DPD at corresponding scale 
$\mu \simeq 1/y$ should be given by eq.~\eqref{split-dpd} (at leading order in $\alpha_s$, which we
restrict ourselves to here). In the unpolarised case we have  $T_{i\to jk}(x) = P_{i \to j}(x)$, where 
$P_{i \to j}(x)$ is the leading-order splitting function appearing in single PDF evolution, without the 
virtual terms proportional to $\delta(1-x)$. At nonperturbative $y \sim 1/\Lambda$, an ansatz is required. 
We use the following form:
\begin{align} \label{eq:Fmodelunpol}
 F^{ij}(x_1, x_2, \tvec{y},\mu_y) &=F^{ij}_{\text{spl}}(x_1, x_2, \tvec{y},\mu_y) + F^{ij}_{\text{int}}(x_1, x_2, \tvec{y},\mu_y)
\\
F^{ij}_{\text{int}}(x_1,x_2, \tvec{y}, \mu_y) &= \frac{1}{4\pi h_{ij}}\, e^{- \tfrac{y^2}{4h_{ij}}}\,
    f_{i}(x_1,\mu_y)\, f_{j}(x_2,\mu_y) (1-x_1-x_2)^2 (1-x_1)^{-2} (1-x_2)^{-2}
\\
F^{ij}_{\text{spl}}(x_1,x_2, \tvec{y}, \mu_y) &= \frac{1}{\pi y^2}\, e^{- \tfrac{y^2}{4h_{ij}}}\;
   \frac{\alpha_s(\mu_y)}{2\pi}\, 
    \sum_k \frac{f_k(x_1+x_2,\mu_y)}{x_1+x_2}\,
   P_{k\to i}\biggl( \frac{x_1}{x_1+x_2} \biggr)
\end{align}
with 
\begin{align} \label{muydef}
\mu_y &= \frac{2e^{-\gamma_E}}{y^*} \equiv \frac{b_0}{y^*} \,, \qquad 
y^{*} = \frac{y}{\sqrt{ 1 + y^2 /y_{\text{max}}^2 }}
\end{align}

$F_{\text{spl}}$ is essentially the contribution to the DPD from perturbative splitting,
whilst $F_{\text{int}}$ represents a contribution to the DPD from parton pairs already
existing at the low scale $\Lambda$. The prescription in eq.~\eqref{muydef} is designed to 
freeze the scale in the PDFs and $\alpha_s$ as $y$ approaches $y_{\text{max}}$, where
$y_{\text{max}}$ is taken of order $1/\Lambda$. This avoids evaluations of the PDFs 
and $\alpha_s$ at very low scale, and is similar to the $b^*$ prescription used in 
TMD phenomenology \cite{Collins:1981va,Collins:1984kg}. Here we take $y_{\text{max}} 
= 0.5 \text{GeV}^{-1}$.

For the non-splitting piece $F^{ij}_{\text{spl}}$ we make the traditional ansatz
of a product of single PDFs, multiplied a smooth function with width in $y$ of 
order of the transverse proton size. Here we additionally multiply by a function of the
$x_i$ that doesn't affect the DPD at small $x_i$, but smoothly cuts it off 
near the kinematic bound $x_1+x_2=1$ -- the function we use is that given in 
eq.~(3.12) of \cite{Gaunt:2009re}, with $n$ set to $2$. For the $y$-dependent 
function, we use a simplified version of the one used in section 4.1 of \cite{Diehl:2014vaa}, 
where we now take the width $h$ to be $x$-independent (corresponding to the $h(x_1,x_2)$ of 
\cite{Diehl:2014vaa} evaluated at $x_1=x_2=10^{-3}$), and we set each $h$ with 
$q^-$ indices to be the same as the one with $q^+$. Then we have:
\begin{align}
 h_{ij} = h_{i} + h_{j}
\end{align}
with
\begin{align}
 h_{g} = 2.33 \text{GeV}^{-2} \qquad h_{q} = h_{\bar{q}} = 3.53 \text{GeV}^{-2}
\end{align}

We include the same Gaussian damping in $F_{spl}$ to ensure the overall $F$ dies off quickly 
to zero at $y$ values much larger than the transverse proton size.

The DPD is evolved from the initial scale $\mu_y$ to final scale $\mu$ using the appropriate renormalisation group equation for the DPDs -- namely the homogeneous double DGLAP equation
(given in, for example, eq.~(5.93) of \cite{Diehl:2011yj}). In practice this is achieved using
a modified version of the code developed in \cite{Gaunt:2009re}.

For the cut-off, a theta function is used for simplicity -- i.e.~$\Phi(\nu y) = \Theta(\nu y - b_0)$. We set $\mu$ in the DPDs to $80$ GeV (appropriate for the production of a $W$ boson pair).
In this investigation, we take the collider energy to be 14 TeV, and set $x_1$ and $\bar{x}_1$ to correspond to the central production of a $W$ boson, with $x_2$ and $\bar{x}_2$ corresponding to the production of a $W$ boson with rapidity $Y_1$:
\begin{align}
x_1 = \bar{x}_1 = 5.7\times 10^{-3} \quad x_2 = 5.7\times 10^{-3} \exp(Y_1) \quad \bar{x}_2 = 5.7\times 10^{-3} \exp(-Y_1)
\end{align}

\begin{figure}
\begin{center}
\subfigure[]{\includegraphics[height=10.5em]{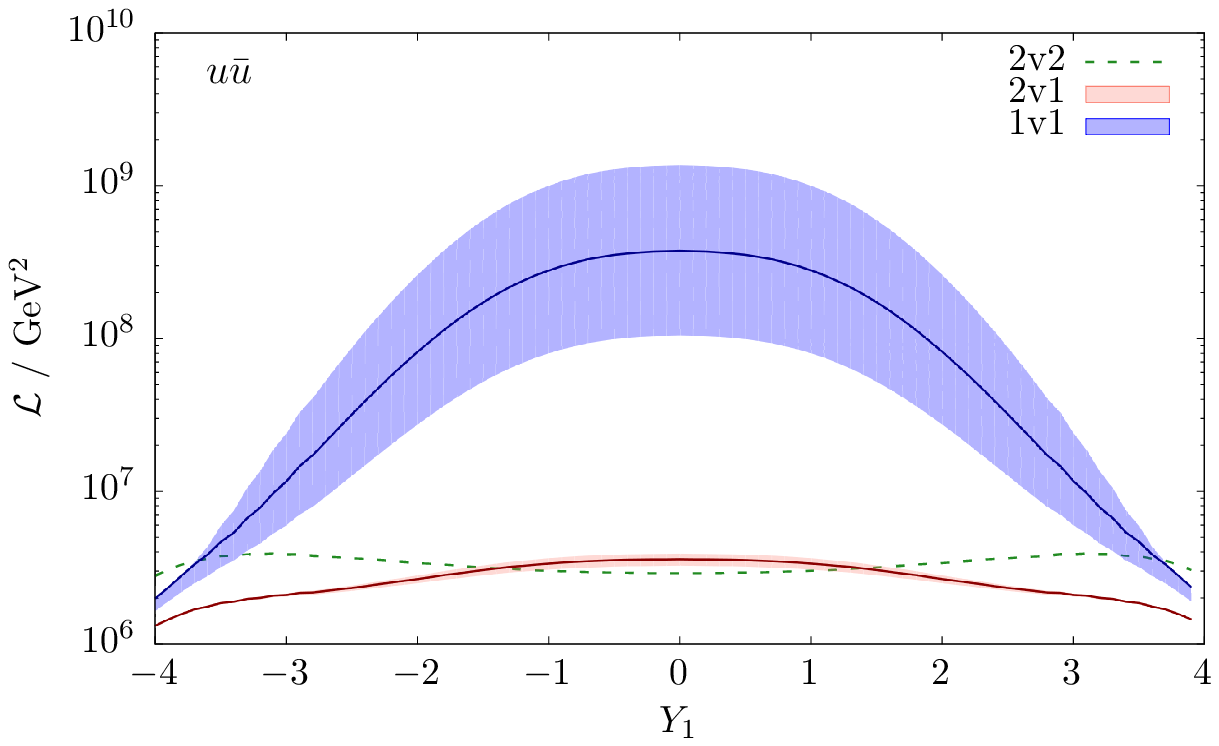}}
\hspace{1em}
\subfigure[]{\includegraphics[height=10.5em]{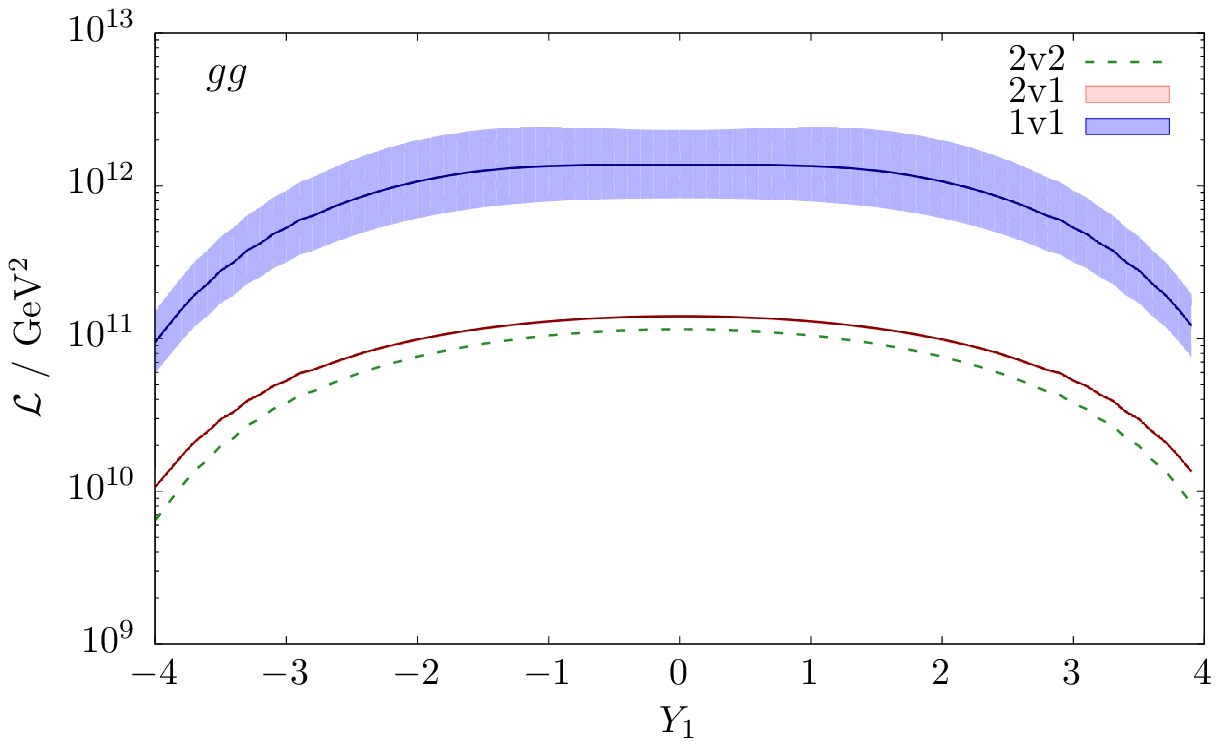}}
\hspace{1em}
\subfigure[]{\includegraphics[height=10.5em]{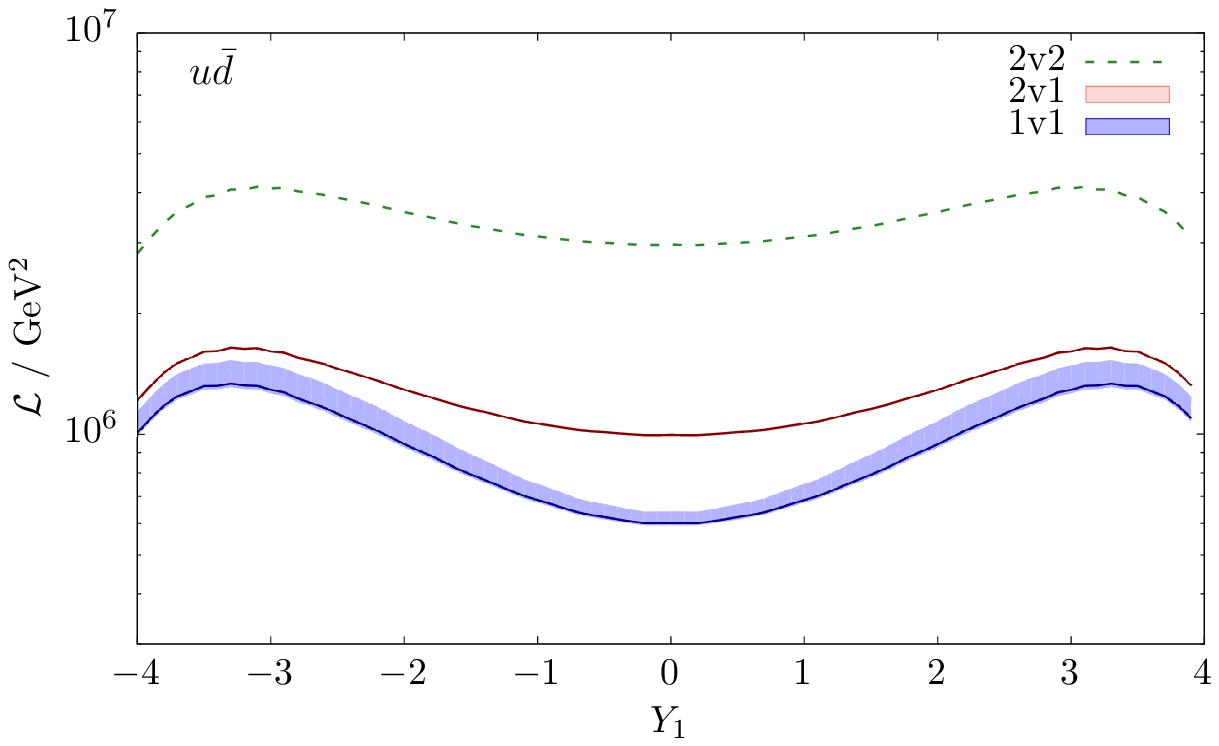}}
\caption{\label{fig:DPSlumi} Double parton scattering luminosities $\mathcal{L}^{ijkl}(Y_1)$
for the production of two systems with scale $\mu =80$ GeV in a $14$ TeV collider, one with
central rapidity, and the other with rapidity $Y_1$. Figure (a) corresponds to 
$ijkl = u\bar{u}\bar{u}u + \bar{u}uu\bar{u}$, (b) corresponds to $ijkl = gggg$,
and (c) to $ijkl = u\bar{d}\bar{d}u + \bar{d}uu\bar{d}$.}
\end{center}
\end{figure}

In figure \ref{fig:DPSlumi}, we plot $\mathcal{L}^{ijkl}(Y_1)$ in the range $-4 \le Y_1 \le 4$ 
for the parton combinations $ijkl = u\bar{u}\bar{u}u + \bar{u}uu\bar{u}$ (figure \ref{fig:DPSlumi}(a)), 
$ijkl = gggg$ (figure \ref{fig:DPSlumi}(b)), and $ijkl = u\bar{d}\bar{d}u + \bar{d}uu\bar{d}$ (figure \ref{fig:DPSlumi}(c)). 
The first parton combination appears in e.g. $ZZ$ production, the second is important in four-jet production, and the 
last appears in $W^+W^+$. We split the overall luminosity into a 1v1 contribution ($F_{\text{spl}} \otimes F_{\text{spl}}$), 2v1 contribution
($F_{\text{spl}} \otimes F_{\text{int}} + F_{\text{int}} \otimes F_{\text{spl}}$) and 2v2 
contribution ($F_{\text{int}} \otimes F_{\text{int}}$). We also vary $\nu$ by a factor of $2$ 
around a central value of $80$ GeV in each contribution to show how the DPS contribution alone
is affected by variation of this cutoff. The bands in each figure are generated using the extremal
values of $\nu$, whilst the line denotes the luminosity with $\nu = 80$ GeV.

We immediately notice in figures \ref{fig:DPSlumi}(a) and (b) that the 1v1 contribution is generally much larger than
the 2v2 and 2v1 contributions, with enormous $\nu$ variation in this former piece. This shows
that for these channels, and for these scales and $x$ values, that one must include the SPS 
corrections up to the order that includes figure \ref{fig:split}(b) together with the subtraction,
so as to cancel the $\nu$ dependence and obtain a sensible prediction. By contrast, in
figure \ref{fig:DPSlumi}(c) the 1v1 contribution is small compared to the 2v1 and 2v2, with small $\nu$ dependence.
This is because, as opposed to $u\bar{u}$ and $gg$, there is no leading-order splitting directly 
giving $u\bar{d}$ (generation of a $u\bar{d}$ pair requires at least two steps, such as $u \to 
u + g \to  u + d + \bar{d}$). Here, there is less of a need to compute the SPS term up to the order
that contains the first nonzero DPS-type loop (in both amplitude and conjugate), and corresponding
subtraction, to compensate the $\nu$ dependence. This is fortunate, since in this case one would 
require an SPS calculation two orders higher than that of figure \ref{fig:split}(b) (two-step rather than
one-step splittings are required in both protons), which is well beyond the current state of
the art.

\section*{Acknowledgements}

J.G.\ acknowledges financial support from the European Community under the
Ideas program QWORK (contract 320389).

\bibliographystyle{jhep}
\bibliography{jgauntqcd16}

\end{document}